\documentclass[12pt,preprint]{aastex}

\usepackage{indentfirst}
\usepackage{floatflt,graphicx,rotating}
\usepackage{longtable}
\usepackage{amsmath}
\usepackage{lscape}
\newcommand{\nulx}{N$_{ULX}(SFR)$}
\newcommand{\chandra}{{\em Chandra}}
\newcommand{\cxo}{{\em CXO}}
\newcommand{\wave}{{\sc Wavdetect}}

\shorttitle{Ultra-Luminous X-Ray Sources in the Most Metal Poor Galaxies}
\shortauthors{Prestwich et al.}

\begin{document}

\title{Ultra-Luminous X-Ray Sources in the Most Metal Poor Galaxies}

\author{A.H. Prestwich}
\affil{Harvard-Smithsonian Center for Astrophysics, 60 Garden Street, Cambridge, MA 02138}
\author{Maria Tsantaki}
\affil{University of Crete, Physics Department, 71003 Heraklion, Crete, Greece \& Centro de Astrofsica, Universidade do Porto,
Rua das Estrelas, 4150-762 Porto, Portugal}
\author{A. Zezas}
\affil{Harvard-Smithsonian Center for Astrophysics, 60 Garden Street, Cambridge, MA 02138 \& University of Crete, Physics Department, 71003 Heraklion, Crete, Greece \& Foundation for Research and Technology-Hellas, 71110 Heraklion, Crete, Greece}
\author{F. Jackson}
\affil{Harvard-Smithsonian Center for Astrophysics, 60 Garden Street, Cambridge, MA 02138}
\author{T.P. Roberts}
\affil{Department of Physics, University of Durham, South Road, Durham DH1 3LE, UK}
\author{R. Foltz}
\affil{Department of Physics and Astronomy, University of California, Riverside, 900 University Avenue, Riverside, CA 92521}
\author{T. Linden}
\affil{University of Santa Cruz, Department of Physics 211 Interdisciplinary Sciences Building, 1156 High Street, Santa Cruz, CA 95064, USA}
\author{V. Kalogera}
\affil{Center for Interdisciplinary Exploration and Research in Astrophysics (CIERA) \& Department of Physics and Astronomy, Northwestern University, 2145 Sheridan Road, Evanston, IL 60208, USA}

%%%%%%%%%%%%%%%%%%%%%%%%%%%%%%%%%%%% FRONT PAGE %%%%%%%%%%%%%%%%%%%%%%%%%%%%%%%%%%

\begin{abstract}
Ultra-Luminous X-ray sources (ULX) are X-ray binaries with L$_x$ $>10^{39}$ ergs s$^{-1}$.   The most spectacular examples of ULX occur in starburst galaxies and are now understood to be young, luminous High Mass X-ray Binaries.     The conditions under which ULX form are poorly understood, but recent evidence suggests they may be more common in low metallicity systems.     Here we investigate the hypothesis that ULX form preferentially in low metallicity galaxies by searching for ULX in a sample of Extremely Metal Poor Galaxies (XMPG) observed with the \chandra\ X-ray Observatory (\cxo).   XMPG  are defined as galaxies with log$(O/H)+12<7.65$, or less than 5\% solar.     These are the most metal-deficient galaxies known, and a logical place to find ULX if they favor metal poor systems.    We compare the number of ULX (corrected for background contamination) per unit of star formation (\nulx) in the XMPG sample with \nulx\ in a comparison sample of galaxies with higher metallicities  taken from the Spitzer  Infrared Galaxy Sample (SINGS).      We find that ULX occur preferentially in the metal poor sample with a formal statistical significance of  2.3$\sigma$.     We do not see strong evidence for a trend in the formation of ULX in the high metallicity sample:  above 12+log(O/H)$\sim$8.0 the efficiency of ULX production appears to be flat.  The effect we see is strongest in the lowest metallicity bin.  We discuss briefly the implications of these results for the formation of black holes in low metallicity gas.

\end{abstract}

\section{Introduction}

Ultra-Luminous X-ray sources (ULX) are X-ray binaries with L$_x$ $>10^{39}$ ergs s$^{-1}$ most commonly found in star forming and starburst galaxies.     These sources  have attracted
considerable attention in recent years because they have broad-band
X-ray luminosities many times the Eddington limit for a neutron star
or stellar mass black hole \citep{Miller2004,Roberts2007,Feng2011}.   Some of these extreme objects may be intermediate mass black holes (M$>$500M$_{\odot}$,  \citep{Colbert1999, Farrell2009,Sutton2012} .   It seems likely, however that most ULX are stellar X-ray sources which are either radiating in excess of the Eddington limit and/or have black-hole  masses somewhat higher than is commonly seen in black hole candidates in Milky Way binaries  (M$>$10$_{\odot}$,   \citet{Roberts2007, Gladstone2009, Zampieri2009} ).      ULX are found in a wide variety of systems -- spirals, interacting starbursts and dwarf galaxies.   However, the conditions under which ULX form are poorly understood.

It is well established that the number of ULX in star forming galaxies scales with the star formation rate \citep{Grimm2003, Ranalli2003,Mapelli2010,Mineo2012}. In addition, there are are several lines of evidence to suggest that ULX form preferentially in low metallicity gas.   \cite{Kaaret2011} found that the ratio of  X-ray luminosity to star formation rate is an order of magnitude larger in low metallicity Blue Compact Dwarf Galaxies than for solar metallicity star forming galaxies.   \citet{Swartz2008} found that the occurrence rate per unit galaxy mass is higher in dwarfs  than in more massive galaxies (see also \citet{Walton2011}).    This surprising correlation is explained if ULX favor the metal poor environments found in  dwarfs.   
Evidence that there is a direct connection between metallicity and ULX production was presented by \citet{Mapelli2010}  who find an anti-correlation between the number of ULX and metallicity based on a sample of 64 galaxies.  A further study by \citet{Mapelli2011} with the addition of two XMPG came to the same conclusion.
Spectroscopy of ULX counterparts and surrounding gas also suggest that ULX are formed from stars in metal poor gas  (e.g. \citet{Soria2005, Liu2007}).    
 
   Here we investigate the hypothesis that ULX form preferentially in low metallicity galaxies by searching for ULX in a sample of Extremely Metal Poor Galaxies (XMPG) observed with the \chandra\ X-ray Observatory (\cxo).   XMPG  are defined as galaxies with log$(O/H)+12<7.65$, or less than 5\% solar \citep{Papaderos2008}.    These are the most metal-deficient galaxies known, and a logical place to find ULX if they favor metal poor systems.    Our goal is to compare the number of ULX (normalized to the star formation rate and accounting for cosmic background sources) in a sample of XMPG with a comparison sample of  galaxies with higher metallicities  taken from the Spitzer Infrared Galaxy Sample (SINGS).   For a sample of galaxies we define  \nulx\ to be:
  \begin{equation}
  N_{ULX}(SFR)=\frac {\sum N_{ULX}-\sum N_{BKG}}{\sum SFR}
 \end{equation}
 
 Here $\sum N_{ULX}$ is the total number of ULX found in a sample of galaxies (e.g. the XMPG sample), $\sum N_{BKG}$ the total number of expected cosmic background sources (with an apparent luminosity $> 10^{39}$ ergs s$^{-1}$) in the same sample and $\sum SFR$  is the integrated star formation rate (in M${_\odot}$ per year) of the sample galaxies.   
  
\section{Galaxy sample}
\label{galaxy_sample}

Our sample  is listed in Table~\ref{tab:xmpg_sample}.   It consists of 25 nearby (d$\leq$50 Mpc) extremely metal poor galaxies ($(O/H)+12<7.65$).  These are the most metal-deficient galaxies known. Most of those are Blue Compact Dwarf galaxies.  Three galaxies in the sample   (I Zw18, SBS 0335-052, SBS 0335-052W) were observed with Chandra in 2000 \citep{Thuan2004}  The remainder were observed as part of a Cycle 11 Large Project.  The exposure times were set to obtain a 3.5$\sigma$ detection of a point source of luminosity  7.8$\times 10^{38}$ ergs cm$^{-2}$ s$^{-1}$ ( i.e.  our survey is complete down to 7.8$\times 10^{38}$ ergs cm$^{-2}$ s$^{-1}$.)   Completeness is further discussed in Section~\ref{Completeness}.   Observational details are given in Table~\ref{tab:xmpg_obs}.

%%%%%%%%%%%%%%%%%%%%%%%%%%%%%%%%  table 2 %%%%%%%%%%%%%%%%%%%%%%%%%%%%%%%%%%%%%%%%%%%%%%%%%%%%%%%%%%%%%%%%%%%%%%
\begin{table}
 \caption{The Extremely Metal Poor Galaxy sample.}
 \makebox[\linewidth]{
 \footnotesize
  \begin{tabular}{l c c c c c c}
   \hline\hline
    Galaxy & R.A. & Dec. & Distance & D$_{25}$ & N$_{H}$ & 12+log(O/H) \\ [0.5ex]
           & (J2000) & (J2000) & (Mpc) & (arcmin) & (10$^{20}$ cm$^{-2}$) &      \\
    \hline
UGC 772 & 08h25m55.5s &	+35d32m32s & 3.10 & 1.20$\times$0.90 & 11.5 & 7.24$\pm$0.03 \\
SDSS J210455.31-003522.2 & 21h04m55.3s & -00d35m22s & 13.7 & - & 6.47 & 7.26$\pm$0.03 \\
SBS 1129+576 &	11h32m02.5s & +57d22m46s & 26.3 & 0.75$\times$0.10 & 0.87 & 7.41$\pm$0.07 \\
HS 0822+3542 & 08d25m55.5s & +35d32m32s & 12.7 & 0.27$\times$0.12 & 4.82 & 7.35 \\
SDSS J120122.32+021108.5 & 12h01m22.3s & +02d11m08s & 18.4 & 0.35$\times$0.12 & 1.88 & 7.55$\pm$0.03 \\
RC2 A1116+51 & 11h19m34.3s & +51d30m12s & 20.8 & 0.24$\times$0.17 & 1.19 & 7.51$\pm$0.04 \\
SBS 0940+544 &	09h44m16.6s & +54d11m34s & 24.7 & 1.22$\times$1.54 & 1.34 & 7.48 \\
KUG 1013+381 & 	10h16m24.5s & +37d54m46s & 19.6 & 0.38$\times$0.25 & 1.41 & 7.58 \\
SBS 1415+437 & 14h17m01.4s & +43d30m05s	& 10.4 & 0.75$\times$0.15 & 1.21 & 7.60 \\
6dF J0405204-364859 & 04h05m20.3s & -36d49m01s & 11.0 & 0.48$\times$0.35 & 0.88 & 7.34\\	
SDSS J141454.13-020822.9 & 14h14m54.1s & -02d08m23s & 24.6 & 0.34$\times$0.24 & 4.17 & 7.32 \\
SDSS J223036.79-000636.9 & 22h30m36.8s & -00d06m37s & 18.0 & 0.24$\times$0.20 & 5.20 & 7.64 \\
UGCA 292 & 12h38m40.0s & +32d46m01s & 3.5 & 1.00$\times$0.70 & 1.34 & 7.27$\pm$0.08 \\
HS 1442+4250 & 14h44m12.8s & +42d37m44s & 10.5 & 1.13$\times$0.26 & 1.53 & 7.63 \\
KUG 0201-103 &	02h04m25.6s & -10d09m35s & 22.7	 & 0.46$\times$0.19 & 2.08 & 7.56 \\
SDSS J081239.52+483645.3 & 08h12m39.5s & +48d36m45s & 9.04 & 0.46$\times$0.23 & 4.58 & 7.16 \\
SDSS J085946.92+392305.6 & 08h59m46.9s & +39d23m06s & 10.9 & 0.39$\times$0.27 & 2.44 & 7.45 \\
KUG 0743+513 & 07h47m32.1s & +51d11m28s & 8.6 & 0.70$\times$0.30 & 5.17 & 7.68 \\
KUG 0937+298 & 09h40m12.8s & +29d35m30s & 11.2 & 0.62$\times$0.23 & 1.87 & 7.45 \\
KUG 0942+551 & 09h46m22.8s & +54d52m08s & 24.4 & 0.37$\times$0.18 & 1.23 & 7.66 \\
SBS 1102+606 & 11h05m53.7s & +60d22m29s	& 19.9 & 0.60$\times$0.28 & 0.59 & 7.64$\pm$0.04 \\
RC2 A1228+12 & 12h30m48.5s & +12d02m42s	& 21.2 & 0.29$\times$0.16 & 2.47 & 7.64 \\
I Zw 18 & 09h34m02.0s & +55d14m28s & 17.1 & 0.30$\times$0.20 & 1.99 & 7.17 \\
SBS 0335-052 & 03h37m44.0s & -05d02m40s & 52.6 & 0.23$\times$0.20 & 4.98 & 7.25$\pm$0.05 \\
SBS 0335-052W & 03h37m38.4$\times$s & -05d02m37s & 52.2 & 0.10$\times$0.13 & 4.96 & 7.10$\pm$0.08 \\
    \hline\hline
    \end{tabular} } 
\footnotesize
\vskip 0.3cm
NOTES: Distances are taken from the NASA/IPAC Extragalactic Database (NED). D$_{25}$ shows the diameter of major$\times$minor axis. The Galactic $N_{H}$ is calculated with Colden tool, http://cxc.harvard.edu/toolkit/colden.jsp. Metallicities are taken from literature. For the galaxy SDSS J210455.31-003522.2 we used the optical diameter from NASA/SAO Image Archive.
\label{tab:xmpg_sample}
\end{table} 
%%%%%%%%%%%%%%%%%%%%%%%%%%%%%%%%%%%%%%%%%%%%%%%%%%%%%%%%%%%%%%%%%%%%%%%%%%%%%%%%%%%%%%%%%%%%%%%%%%%%%%%%%%%%%%%%%

%%%%%%%%%%%%%%%%%%%%%%%%%%%%%%%%  table 3    %%%%%%%%%%%%%%%%%%%%%%%%%%%%%%%%%%%%%%%%%%%%%%%%%%%%%%%%%%%%%%%%%%%%%%
\begin{table}[!h]
\caption{X-ray observations}
\makebox[\linewidth]{
 \begin{tabular}{l c c c c}
 \hline\hline
Galaxy & Obs.ID & Date & Exposure (sec) & Instrument \\ [0.5ex]
\hline
UGC 772 & 11281 & 30-08-2009 & 5081 & ACIS-S3 \\
SDSS J210455.31-003522.2 & 11282 & 04-09-2009 & 5007 & ACIS-S3 \\
SBS 1129+576 & 11283 & 06-07-2010 & 14755 & ACIS-S3 \\
HS 0822+3542 & 11284 & 20-12-2009 & 51200 & ACIS-S3 \\
SDSS J120122.32+021108.5 & 11286 & 23-11-2009 & 8097 & ACIS-S3 \\
RC2 A1116+51 & 11287 & 07-11-2009 & 11640 & ACIS-S3 \\
SBS 0940+544 & 11288 & 18-01-2010 & 16828 & ACIS-S3 \\
KUG 1013+381 & 11289 & 24-01-2010 & 9402 & ACIS-S3 \\
SBS 1415+437 & 11291 & 30-102009 & 5114 & ACIS-S3 \\
6dF J0405204-364859 & 11292 & 28-05-2010 & 5010 & ACIS-S3 \\	
SDSS J141454.13-020822.9 & 11293 & 18-12-2009 & 16680 & ACIS-S3 \\
SDSS J223036.79-000636.9 & 11294 & 25-09-2009 & 7715 & ACIS-S3 \\
UGCA 292 & 11295 & 06-11-2009 & 5007 & ACIS-S3 \\
HS 1442+4250 & 11296 & 26-11-2009 & 5188 & ACIS-S3 \\
KUG 0201-103 & 11297 & 06-09-2009 & 13590 & ACIS-S3 \\
SDSS J081239.52+483645.3 & 11298 & 18-12-2009 & 4777 & ACIS-S3 \\
SDSS J085946.92+392305.6 & 11299 & 18-12-2009 & 4782 & ACIS-S3 \\
KUG 0743+513 & 11300 & 18-12-2009 & 5073 & ACIS-S3 \\
KUG 0937+298 & 11301 & 16-01-2010 & 5007 & ACIS-S3 \\
KUG 0942+551 & 11302 & 19-01-2010 & 16020 & ACIS-S3 \\
SBS 1102+606 & 11285 & 23-08-2010 & 10340 & ACIS-S3 \\
RC2 A1228+12 & 11290 & 26-07-2010 & 12200 & ACIS-S3 \\
I Zw 18 & 805 & 08-02-2000 & 40750 & ACIS-S3 \\
SBS 0335-052 & 796 & 07-09-2000 & 59742 & ACIS-I3 \\
SBS 0335-052W & 796 & 07-09-2000 & 59742 & ACIS-I3 \\
\hline\hline
 \end{tabular} }
\label{tab:xmpg_obs}
\end{table} 
%%%%%%%%%%%%%%%%%%%%%%%%%%%%%%%%%%%%%%%%%%%%%%%%%%%%%%%%%%%%%%%%%%%%%%%%%%%%%%%%%%%%%%%%%%%%%%%%%%%%%%%%%%%%%%%%%

\section{X-Ray Obervations}
\label{xray_obs}

  All observations were obtained with the back-illuminated chip ACIS-S3 except for the galaxies SBS 0335-052 and SBS 0335-052W, which were observed with the front-illuminated ACIS-I3 camera. The observations were performed in VFAINT mode.   We reprocessed raw data (level 1 event files) using Chandra Interactive Analysis of Observations software (CIAO), version 4.2 and the CALibration DataBase (CALDB), version 4.3.0.   Standard routines were used to correct for bad pixels, charge transfer inefficiency and time dependent gain.    A new Level 2 events file was created by filtering for standard grades (0, 2, 3, 4, 6) and rejecting grades associated with bad pixels.

 The positions of X-ray sources were determined with the CIAO tool \wave. This tool is a wavelet-based source detection algorithm \citep{Freeman2002}.    The X-ray image is convolved with a wavelet function to produce a "correlation image".   Clumps of counts (sources) are identified as a local maximum in the correlation image if the scale of the wavelet is approximately equal to (or greater than) the size of the clump.   Hence, \wave\ also gives an estimate of the size of the source.    It is typically run with wavelets of differing scales to better detect extended emission.  
 
   \wave\ was run on the level 2 files in the 0.3-8keV band using scales of  1.0, 2.0, 4.0, 8.0 and 16.0 pixels. We set a threshold significance for identifying a pixel as belonging to the source at 10$^{-6}$.    All the detected sources have scales consistent with their being point sources.   Sources within the D$_{25}$ region (defined as the elliptical contour best corresponding to the 25 mag arcsec$^{-2}$ blue isophote \citet{deVaucouleurs1991}) were considered to be associated with the galaxy. The results of the source detection are show in Table~\ref{tab:xmpg_xsources}.

%%%%%%%%%%%%%%%%%%%%%%%%%%%%%%%%  table 5    %%%%%%%%%%%%%%%%%%%%%%%%%%%%%%%%%%%%%%%%%%%%%%%%%%%%%%%%%%%%%%%%%%%%%%
\begin{table}[!h]
\caption{X-ray source detection.}
\makebox[\linewidth]{
\begin{tabular}{l c c}
\hline\hline
Galaxy & Number of sources & Position \\ 
 &  & R.A. (J2000) Dec. (J2000)\\ [0.5ex]
\hline
SBS 1129+576 & 1 & 11h32m02.57s +57d22m36.87s \\
RC2 A1116+51 &  1 & 11h19m34.13s +51d30m12.48s \\
SBS 0940+544 & 1 & 09h44m16.44s +54d11m34.39s \\
I ZW 18 & 1 & 09h34m01.97s +55d14m28.23s \\
SBS 0335-052 & 1 & 03h37m44.04s -05d02m39.89s \\
SBS 0335-052W & 1 & 03h37m38.44s -05d02m37.14s \\
\hline\hline
\end{tabular} }
\label{tab:xmpg_xsources}
\end{table}
%%%%%%%%%%%%%%%%%%%%%%%%%%%%%%%%%%%%%%%%%%%%%%%%%%%%%%%%%%%%%%%%%%%%%%%%%%%%%%%%%%%%%%%%%%%%%%%%%%%%%%%%%%%%%%%%

Source regions were defined  based on the \wave\  source extent.   We also create the corresponding background regions by setting apertures with larger radii than the sources. We perform photometry for these sources, using the {\tt dmextract} tool. The results of the photometry are shown at Table~\ref{tab:xmpg_xphot}. The source counts per pixel follow the Poisson distribution and the count error, since the number of counts is very low, the Gehrels approximation: error = 1 + $\sqrt{(N + 0.75)}$ \citep{Gehrels1986}. 
 
The flux of an X-ray source is proportional to the net count rate, where the constant of proportionality depends on the response of the detector and the assumed source spectrum:

\begin{equation}
 A=\frac{Flux}{Count Rate}, [(erg/sec/cm^{2})/(counts/sec)]
\end{equation}

We extracted   standard spectral responses  (the Redistribution Matrix (RMF) and Area Response Matrix) and simulated a spectrum for each source.      Using the  results of \citet{Swartz2004}, we adopt an  intrinsic source spectrum  with photon index $\Gamma$=1.7 and assume values for galactic absorption taken from \citet{Dickey1990} . We then estimate the constant A from the ratio of the number of counts to the calculated flux in the simulated spectrum.      Fluxes and luminosities  were constructed in the 0.3-8 keV band, using the constant A derived from the simulated spectra.    They are shown in Table~\ref{tab:xmpg_xphot}.   One source, IZw18, had enough counts to perform a spectral fit.   The best fit is consistent with the analysis of  \cite{Thuan2004}  and is shown in Table~\ref{tab:fake}.  The fluxes and luminosities of the other sources are shown in Table~\ref{tab:xmpg_xphot}

\begin{table}[!h]
\caption{Spectral fit to I ZW 18.}
\makebox[\linewidth]{
\small
\begin{tabular}{l c c c c c}
\hline\hline
Galaxy & Source count-rate & $\Gamma$ & N$_{H}$ & $\chi^{2}$/dof \\ 
       & (counts/sec) &  & (10$^{20}$ cm$^{-2}$) & \\ 
\hline 
I ZW 18 &  0.0111 & 1.88$^{+0.22}_{-0.20}$ & 7.54$^{+4.76}_{-6.17}$ & 12.69/19  \\ [0.5ex]
\hline\hline
\end{tabular} } 
\label{tab:fake}
\end{table}

%%%%%%%%%%%%%%%%%%%%%%%%%%%%%%%%  table 6    %%%%%%%%%%%%%%%%%%%%%%%%%%%%%%%%%%%%%%%%%%%%%%%%%%%%%%%%%%%%%%%%%%%%%%
\begin{table}[!h]
\caption{X-ray Photometry}
\makebox[\linewidth]{
\begin{tabular}{lccccc}
\hline\hline
Galaxy & Net counts & Background counts & N$_H$ & Flux & L$_x$\\ 
\hline
& & & (10$^{20}F$ cm$^{-2}$) & (10$^{-15}$erg/cm$^{2}$/sec) & (10$^{39}$erg/sec)\\
SBS 1129+576 & 21.47$\pm$5.71 & 0.19$\pm$1.97 & 0.87 & 1.14$\pm$0.3 & 0.72$\pm$0.19 \\
RC2 A1116+51 & 89.04$\pm$10.48 & 0.21$\pm$1.98 & 1.19 & 5.95$\pm$0.70 & 2.93$\pm$0.35 \\ 
SBS 0940+544 & 39.20$\pm$7.32 & 0.20$\pm$1.97 & 1.34 & 1.81$\pm$0.34 &1.26$\pm$0.24\\
I ZW 18 & 494.44$\pm$23.25 & 1.56$\pm$2.51 &7.54 & 7.46$\pm$0.35 & 4.92$\pm$0.25\\
SBS 0335-052 & 28.21$\pm$6.38 & 1.30$\pm$2.43 & 4.98 &0.56$\pm$0.13 &1.86$\pm$0.42\\
SBS 0335-052W & 116.29$\pm$11.82 & 1.30$\pm$2.43 & 4.96 & 2.31$\pm$0.23 &7.53$\pm$0.77\\
\hline\hline
\end{tabular} }
\label{tab:xmpg_xphot}
\end{table}
  
\subsection{Completeness and Background Sources}
\label{Completeness}

 In this section we demonstrate that our sample of ULX in the metal-poor galaxies is complete: i.e. that we are not "missing"  ULX  due to inadequate exposure.   In addition, we estimate the number of ULX which are chance coincidences:  background AGN that happen to align with the optical galaxy.
 
 The completeness limit is expressed by the source detection probability of a galaxy .   It is a function of the source and background intensity, measured in counts.   The detection probability as a function of source counts and background counts/pixel can be parameterized by the following function \citep{Zezas2007}:
\begin{equation}
A(C)=1.0-\lambda_{0}C^{-\lambda_{1}}e^{-\lambda_{2}C} 
\end{equation}
where C is the source intensity in counts, $\lambda_{0}$, $\lambda_{1}$, $\lambda_{2}$ parameters that depend on the background counts per pixel.

 All galaxies of our sample were measured to have background level below 0.025 counts per pixel. The best-fit parameters for this background give a detection probability of the form: 

\begin{equation}
A(C)=1.0-11.12C^{-0.83}e^{-0.43C}.
\end{equation}

 Therefore, by solving the equation for 90\% and 50\% completeness, the corresponding source counts are:
$C_{90}=7.2$ counts and $C_{50}=4.4$ counts.

 Table~\ref{tab:xmpg_xcomp} shows that all luminosities of 90\% completeness and all luminosities 50\% completeness, are well under the limit of 10$^{39}$ ergs s${-1}$ cm$^{2}$.  This means that we have detected at least 90\% of all existing ULX sources in the galaxy sample.   

% %%%%%%%%%%%%%%%%%%%%%%%%%%%%%%%%  table 9    %%%%%%%%%%%%%%%%%%%%%%%%%%%%%%%%%%%%%%%%%%%%%%%%%%%%%%%%%%%%%%%%%%%%%%
 \begin{table}
 \caption{Completeness}
 \makebox[\linewidth]{
 \small
 \begin{tabular}{l c c c c }
 \hline\hline
 Galaxy & Exposure time & 90\% & 50\%  & N$_{BKG}$\\ [0.5ex]
        & (sec) & (10$^{38}$ erg/sec) & (10$^{38}$ erg/sec) & \\
\hline
 UGC 772 & 5081 & 1.66 & 1.01 &  0.014\\
 SDSS J210455.31-003522.2 & 5007 & 2.39 & 1.46 & 0.039 \\
 SBS 1129+576 & 14755 & 2.40 & 1.47 & 0.034\\
 HS 0822+3542 & 5120 & 2.00 & 1.23  & 0.004\\
 SDSS J120122.32+021108.5 & 8097 & 2.66 & 1.63 & 0.010\\
 RC2 A1116+51 & 11640 & 2.37 & 1.45 & 0.012  \\
 SBS 0940+544 & 16828 & 2.31 & 1.41  & 0.776 \\
 KUG 1013+381 & 9402 & 2.60 & 1.59  & 0.026 \\
 SBS 1415+437 & 5114 & 1.35 & 0.82 & 0.010  \\
 6dF J0405204-364859 & 5010  & 1.54 & 0.94 &0.017  \\
 SDSS J141454.13-020822.9 & 16680 & 2.31 & 1.41 & 0.034\\
 SDSS J223036.79-000636.9 & 7715 & 2.67 & 1.63  & 0.012\\
 UGCA 292 & 5007 & 0.08 & 0.05 & 0.041 \\ 
 HS 1442+4250 & 5188 & 1.35 & 0.83  & 0.028 \\
 KUG 0201-103 & 13590 & 2.41 & 1.47 & 0.031 \\
 SDSS J081239.52+483645.3 & 4777 & 1.09 & 0.67 &  0.008\\ 
 SDSS J085946.92+392305.6 & 4782 & 1.58 & 0.97 &0.011 \\
 KUG 0743+513 & 5073 & 0.93 & 0.57 & 0.014 \\
 KUG 0937+298 & 5007 & 1.59 & 0.97 & 0.015\\
 KUG 0942+551 & 16020 & 2.36 & 1.45 &  0.027\\
 SBS 1102+606 & 10340 & 2.44 & 1.49 & 0.048\\
 RC2 A1228+12 & 12200 & 2.37 & 1.43 & 0.014 \\
 I ZW 18 & 25956 & 0.72 & 0.44 & 0.013\\
 SBS 0335-052 & 59742 & 4.74 & 2.89 & 0.070\\
SBS 0335-052W & 59742 & 4.66 & 2.85  &0.020\\
 \hline\hline
 \end{tabular} }
 \label{tab:xmpg_xcomp}
 \end{table}
 
 The predicted number of background sources for each galaxy was estimated as follows.    For each galaxy, we calculate the 0.3-8 keV flux for a source of 10$^{39}$ ergs s$^{-1}$ and then transform the flux to the 0.5-2  keV band assuming a power law source spectrum with photon index $\Gamma$=1.7 and a foreground N$_H$ listed in Table~\ref{tab:xmpg_sample}.    We then used the LogN-LogS curves of \citet{Giacconi2001} for the 0.5-2 KeV band to estimate the number of background sources per square degree and normalize by the D$_{25}$ values to get the absolute number of background sources predicted to lie within the optical area of the galaxy.      These values are given in Table~\ref{tab:xmpg_xcomp}.   We note that SBS 0940+544 has a predicted background value close to one.
    
 %%%%%%%%%%%%%%%%%%%%%%%%%%%%%%%%%%%%%%%%%%%%%%%%%%%%%%%%%%%%%%%%%%%%%%%%%%%%%%%%%%%%%%%%%%%%%%%%%%%%%%%%%%%%%%%%%

%%%%%%%%%%%%%%%%%%%%%%%% SECTION 3 %%%%%%%%%%%%%%%%%%%%%%%%%%%%%%%%%%%%%%%%%%%%%%%%%%%%

\section{Estimates of the Star Formation Rate in the XMPG Sample}

It is well established that the number of ULX in a galaxy scales with the star formation rate \citep{Grimm2003,Ranalli2003,Mapelli2010,Mineo2012}.    We therefore need reliable estimates of the star formation rate in the XMPG sample in order to determine whether \nulx\  is higher in the metal poor sample than in the SINGS sample.   We use two methods to estimate the star formation rate: the  Far Ultraviolet (FUV) luminosity from {\it GALEX} and the 24$\mu$m luminosity from {\it Spitzer}.

\subsection{GALEX Data}

The FUV emission from star forming regions comes directly from young massive O and B stars.    We use the method from \citet{Hunter2010} to relate the FUV luminosity to the star formation rate in dwarf galaxies:

\begin{equation}
 SFR_{\rm FUV}(M_{\odot}/yr) = 1.27 \times 10^{-28} L_{\rm FUV} (erg s^{-1} Hz^{-1}) 
\end{equation}

We obtained GALEX  FUV images from the GalexView version 1.4.6 catalog.   All galaxies in the XMPG sample were detected with the exception of  SDSS J223036.79-000636.9.  This galaxy is excluded from the rest of the analysis in this paper.   Source counts were extracted using the {\it CIAO} {\sc dmextract} routine. This was done with either circular or elliptical source apertures, depending on the morphology of the galaxy in question, as well as an annular background aperture. Both of these apertures were centered on the source, with the circular apertures having a radius of either 10 or 15 pixels, and the ellipses having a semi-major and semi-minor axis of length 28 and 14 pixels or 15 and 7.5 pixels, respectively, depending on the size of the source. All background annuli had an area 8 times that of their respective source apertures. The fluxes were computed using the relationship $f _{\rm FUV}$ = (1.4 $\times$ 10$^{-15}$) $\times$ count rate\footnote{http://galexgi.gsfc.nasa.gov/docs/galex/FAQ/counts\_background.html}. Note that these fluxes are in units of erg cm$^{-2}$ s$^{-1}$ {\AA}$^{-1}$. We converted these values to units of erg cm$^{-2}$ s$^{-1}$ Hz$^{-1}$ by multiplying them by $\lambda^{2}$/c, where $\lambda$ is the effective wavelength of the FUV detector (= 1528 {\AA}), and c is the speed of light in units of {\AA} s$^{-1}$. With the fluxes in these new units, we proceeded to calculate the $L_{\rm FUV}$ values of the sources, which were subsequently corrected for extinction. We then calculated the SFR$_{\rm FUV}$ estimates of the galaxies using the extinction-corrected luminosities.

\begin{table}
%\begin{sidewaystable}
\begin{center}
\caption{UV measurements and SFRs}
\begin{tabular}{ccccc}
\hline
Galaxy & E(B-V)$^{(1)}$ & Count Rate$^{(2)}$ & $L_{\rm FUV}$$^{(3)}$ & SFR$_{\rm FUV}$$^{(4)}$ \\ 
\hline
        & (mag) & (ct s$^{-1}$) & ($\times$ 10$^{26}$ erg s$^{-1}$ Hz$^{-1}$) & (10$^{-3}$ $M_{\odot}$ yr$^{-1}$) \\ 
\hline
\hline
UGC 772	& 0.028 & 5.55$\pm$0.05 & 0.09$\pm$0.001 & 1.09$\pm$0.01 \\
SDSS J210455.31-003522.2 & 0.066 & 1.21$\pm$0.06 & 0.49$\pm$0.02 & 6.22$\pm$0.29 \\
SBS 1129+576 & 0.013 & 3.48$\pm$0.20 & 3.47$\pm$0.20 & 44.02$\pm$2.52 \\
HS 0822+3542 & 0.047 & 1.08$\pm$0.09 & 0.32$\pm$0.03 & 4.11$\pm$0.33 \\
SDSS J120122.32+021108.5 & 0.024 & 1.20$\pm$0.03 & 0.64$\pm$0.02 & 8.07$\pm$0.20 \\
RC2 A1116+51 & 0.015 & 4.38$\pm$0.21 & 2.58$\pm$0.12 & 32.76$\pm$1.57 \\
SBS 0940+544 & 0.013 & 1.56$\pm$0.11 & 1.37$\pm$0.10 & 17.41$\pm$1.25 \\
KUG 1013+381 & 0.015 & 7.63$\pm$0.26 & 4.24$\pm$0.15 & 53.87$\pm$1.87 \\
SBS 1415+437 & 0.009 & 12.69$\pm$0.30 & 1.93$\pm$0.05 & 24.55$\pm$0.59 \\
6dF J0405204-364859 & 0.006 & 5.29$\pm$0.22 & 0.80$\pm$0.03 & 10.12$\pm$0.43 \\
SDSS J141454.13-020822.9 & 0.058 & 0.63$\pm$0.02 & 0.78$\pm$0.02 & 9.88$\pm$0.30 \\
UGCA 292 & 0.016 & 7.15$\pm$0.18 & 0.13$\pm$0.003 & 1.62$\pm$0.04 \\
HS 1442+4250 & 0.013 & 9.00$\pm$0.22 & 1.42$\pm$0.04 & 18.08$\pm$0.45 \\ 
KUG 0201-103 & 0.021 & 1.30$\pm$0.09 & 1.02$\pm$0.07 & 13.00$\pm$0.87 \\
SDSS J081239.52+483645.3 & 0.051 & 0.86$\pm$0.02 & 0.13$\pm$0.003 & 1.71$\pm$0.04 \\
SDSS J085946.92+392305.6 & 0.026 & 0.71$\pm$0.01 & 0.14$\pm$0.001 & 1.72$\pm$0.02 \\
KUG 0743+513 & 0.070 & 12.58$\pm$0.34 & 1.66$\pm$0.04 & 21.12$\pm$0.57 \\
KUG 0937+298 & 0.018 & 5.21$\pm$0.16 & 0.97$\pm$0.03 & 12.34$\pm$0.37 \\
KUG 0942+551 & 0.012 & 1.48$\pm$0.12 & 1.26$\pm$0.10 & 16.01$\pm$1.27 \\
SBS 1102+606 & 0.006 & 4.19$\pm$0.17 & 2.27$\pm$0.09 & 28.79$\pm$1.14 \\
RC2 A1228+12 & 0.027 & 2.69$\pm$0.04 & 2.03$\pm$0.03 & 25.69$\pm$0.39 \\
I ZW18 & 0.032 & 11.52$\pm$0.08 & 5.60$\pm$0.04 & 71.12$\pm$0.52 \\	
SBS 0335-052 & 0.047 & 5.52$\pm$0.20 & 28.41$\pm$1.00 & 360.86$\pm$12.76 \\
SBS 0335-052W & 0.046 & 0.56$\pm$0.01 & 2.84$\pm$0.04 & 36.07$\pm$0.53 \\
\hline
\end{tabular}

    {Notes: $^{(1)}$ Reddening magnitude of the source, used to
    calculate source extinction; $^{(2)}$ Observed
    background-subtracted source count rate; $^{(3)}$
    Extinction-corrected far-UV luminosity of the source; $^{(4)}$
    Star-formation rate, calculated using the formula from Hunter et
    al. (2010).}

\end{center}
\label{tab:galex}
\end{table}

\subsection{Spitzer Data}
\label{IR_data}

The  24$\mu$m emission in galaxies  comes from single photon transient heating of small grains and can be used as a tracer of recent star formation.   \citet{Calzetti2007} derived the following relation between the star formation rate (SFR) and  24$\mu$m emission from calibrating $H_{II}$ regions in nearby galaxies:

\begin{equation}
                           SFR_{\rm IR}(M_{\odot}/ yr)= 1.31\times10^{-38}[L_{24\mu m} (ergs/sec)]^{0.885} ,
 \end{equation}
where for intermediate luminosity galaxies:

 \hspace{35 mm}                       1$\times$10$^{40}$ $<$ L$_{24\mu m}$ $<$3$\times$10$^{44}$ erg/sec

 We acquired data from the Spitzer space telescope that were obtained with the MIPS instrument in the 24 microns band. We obtained the post-BCD (post Basic Calibrated Data) data for the 9 galaxies in the sample that have available MIPS 24$\mu$m data. The data sets consist of several exposures that are interpolated into one mosaic image. From these images we measure the surface brightness in units of MJy/sr. The post-BCD product mosaic pixel sizes are 2.42$\times$2.45 arcsec$^{2}$ for the 24 microns detector. 

 We perform aperture photometry using funtools. We define an aperture for the source with an ellipse that includes the total flux of the source and an elliptical annulus for the background region as depicted in Figures 15 - 16 for the 9 galaxies. In order to calculate the total flux density of each source, we sum the total flux of the source, subtract the background, on a set of pixels and multiply by the number of steradian per pixel. We convert the surface brightness to the flux density with the following formula:
 \begin{equation}
         f_v=\sum_{pixels}2.45\times2.45 \frac{arc\sec^{2}}{pixel} \times 0.023504 \frac{Flux}{arc\sec^{2}}  [mJy]. 
 \end{equation}
 The flux density is converted to monochromatic flux:
 \begin{equation}
                           F=\frac{c}{\lambda} f_v(\lambda). 
 \end{equation}

 The uncertainty on the flux is calculated from the mosaic variance image by adding in quadrature the uncertainty of all pixels within the source aperture.   The results are listed at Table~\ref{tab:xmpg_ir_sfr}.

%%%%%%%%%%%%%%%%%%%%%%%%%%%%%%%%%%%table 10  %%%%%%%%%%%%%%%%%%%%%%%%%%%%%%%%%%%%%%%%%%%%%%%%%%%%%%%%%%%%%%%%%%%%%%
\begin{table}
\caption{24 microns measurements and SFRs}
\makebox[\linewidth]{
\footnotesize
\begin{tabular}{l c c c c}
\hline\hline
Galaxy & Flux density & Flux & $L_{24}$ & SFR \\ [0.5ex]
       & (mJy) & ($10^{-13}$ erg/sec/$cm^{2}$) & (10$^{40}$ erg/sec) & $(M_{\odot}yr^{-1})$ \\
\hline
HS 0822+3542 & 2.23$\pm$0.09 & 2.78$\pm$0.12 & 0.54$\pm$0.02 & 0.002$\pm$0.001 \\
SBS 0940+544 & 1.60$\pm$0.10 & 2.01$\pm$0.13 & 1.47$\pm$0.09 & 0.005$\pm$0.003 \\
KUG 1013+381 & 14.54$\pm$0.17 & 18.2$\pm$0.21 & 8.37$\pm$0.09 & 0.022$\pm$0.002 \\
SBS 1415+437 & 16.67$\pm$0.15 & 20.8$\pm$0.19 & 2.70$\pm$0.02 & 0.008$\pm$0.001 \\ 
UGCA 292 & $<$0.87$\pm$0.31 & 1.09$\pm$0.34 & 0.08$\pm$0.03 & 0.0003$\pm$0.0010 \\
HS 1442+4250 & 3.23$\pm$0.11 & 4.03$\pm$0.14 & 0.53$\pm$0.02 & 0.002$\pm$0.001 \\
SBS 1102+606 & 1.33$\pm$0.10 & 1.67$\pm$0.11 & 0.79$\pm$0.06 & 0.003$\pm$0.002 \\
I Zw 18 & 4.84$\pm$0.11 & 6.05$\pm$0.14 & 2.12$\pm$0.05 & 0.006$\pm$0.001 \\
SBS 0335-052 & 67.25$\pm$0.20 & 84.10$\pm$0.25 & 278.94$\pm$0.84 & 0.480$\pm$0.013 \\
SBS 0335-052W & 0.14$\pm$0.04 & 0.018$\pm$0.005 & 0.58$\pm$0.17 & 0.002$\pm$0.005 \\
\hline\hline 
\end{tabular} }
\vskip 0.3cm
\footnotesize
NOTES: The SFR errors were calculated with error propagation. For the galaxy UGCA 292 we calculate the upper limits.
\label{tab:xmpg_ir_sfr}
\end{table}
%%%%%%%%%%%%%%%%%%%%%%%%%%%%%%%%%%%%%%%%%%%%%%%%%%%%%%%%%%%%%%%%%%%%%%%%%%%%%%%%%%%%%%%%%%%%%%%%%%%%%%%%%%%%%%%%%

\subsection{Comparison of {\it Spitzer} and {\it GALEX} star formation rates}

\begin{figure}
  \centering
  \includegraphics{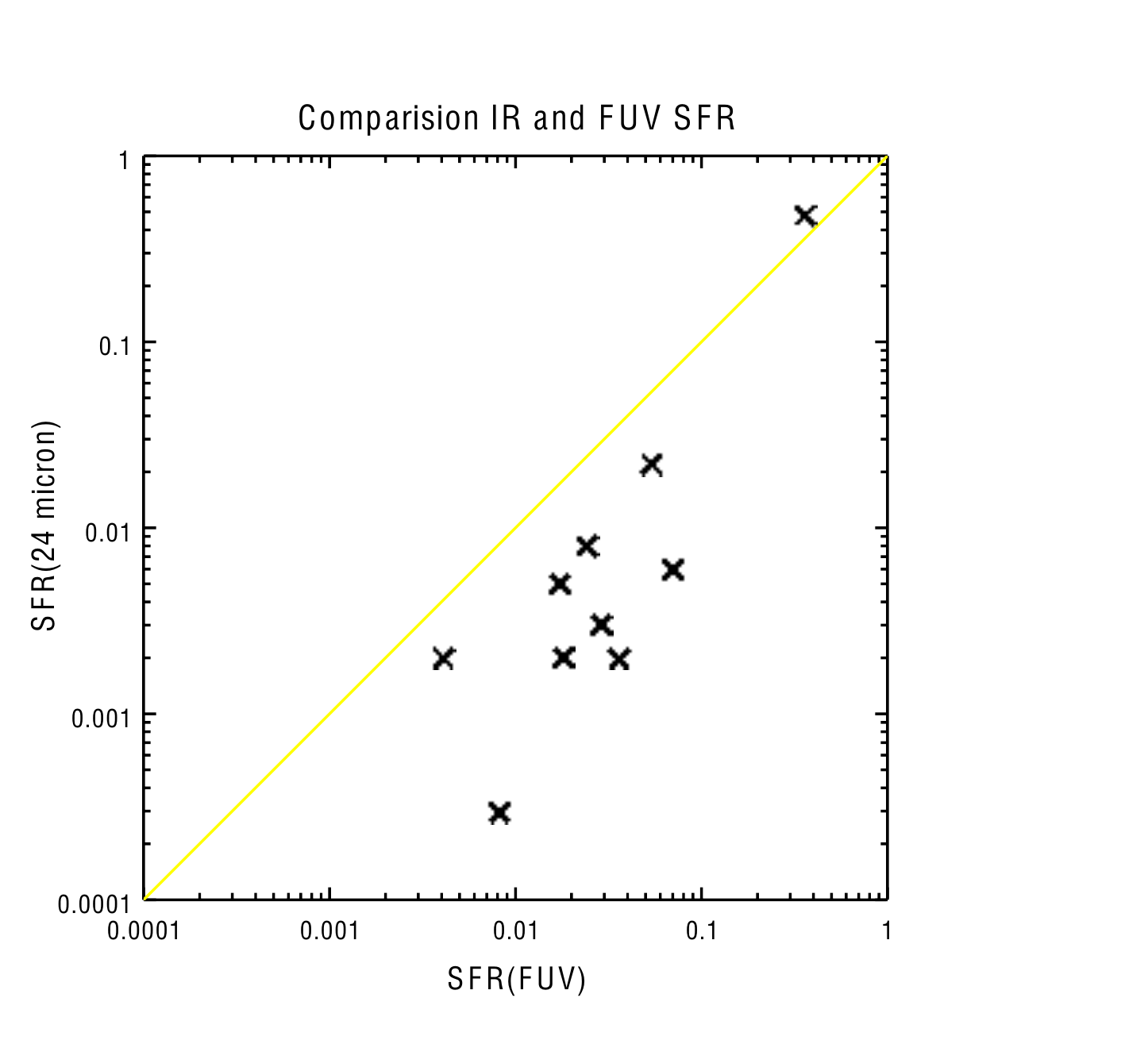}
  \caption{Comparision of the star formation rate determined from the FUV luminosity and 24-$\micron$ luminosity.  The line denotes SFR$_{IR}$=SFR$_{FUV}$. }
  \label{compare_sfr}
\end{figure}

 Several galaxies in the XMPG sample have  both GALEX UV images  and  Spitzer 24 $\mu$m data.  It is therefore instructive to compare SFR derived from the two methods.     Figure~\ref{compare_sfr}   shows the star formation rate derived from the FUV flux with the star formation rate derived from the 24$\micron$ flux.   There is a clear correlation between the two methods.  The rate derived from the FUV flux is systematically higher than for the 24$\micron$ flux.    This is almost certainly because 24$\micron$ emission  underestimates the star formation rate because of the very low dust content of the XMPG.    We adopt the GALEX derived star formation rate because (1) the  Spitzer 24 $\mu$m data likely underestimates the star formation rate, and (2) it is available for all the galaxies.   We note that SBS 0335-052 appears to have very high SFR in comparison with the other XMPGs. 
 
 %----------------------------------- FIGURE-----------------------------------
 %------------------------------------------------------------------------------

\section{The SINGS sample.}
\label{The SINGS sample.}

In order to determine whether there is a statistically significant excess of ULX in the most metal poor systems, we require a comparison sample of galaxies with well determined star formation rates and metallicities.   The Spitzer Infrared Nearby Galaxies Survey \citep{Kennicutt2003} is ideal for this purpose.   It is a comprehensive imaging and spectroscopic study of 75 nearby galaxies with distances $<$ 30 Mpc. The morphological types of the sample range from elliptical to irregular. The SINGS sample does not include absolute extremes in properties that can be found in larger volumes such as ULIRGs, luminous AGNs or extremely metal poor galaxies.   A distance-limited  sample of the SINGS survey has been observed by \chandra\ (Jenkins et al in preparation).  We use a sub-sample of SINGS galaxies  selected by  \citet{Calzetti2007}  which have Hubble Space Telescope (HST) NICMOS images in the Pa$\alpha$ hydrogen emission line (1.8756 $\mu$m),  H$\alpha$ observations and \chandra\ observations. The H$\alpha$ and the Pa$\alpha$ lines are used to measure the extinction correction. Furthermore, the Pa$\alpha$ line is used to calibrate the mid-infrared emission.    There are 33 galaxies in the   \citet{Calzetti2007} sample, 26 of which also form part of the \chandra\ survey.       The galaxies are divided into three groups according to their oxygen abundance: High-metallicity galaxies [12+log(O/H) $\geq$ 8.35], Intermediate-metallicity galaxies [8.00 $<$ 12+log(O/H) $\leq$ 8.35], and Low-metallicity galaxies [12+log(O/H) $\leq$ 8.00]. For reference, the XMPGs have metallicities 12+log(O/H) $\leq$ 7.65.   We list the main characteristics of the SINGS sample in  Table~\ref{tab:sings_sample}.  

The metallicities given in Table~\ref{tab:sings_sample} are given as an upper and lower bound and taken from \citet{Moustakas2010}.  The lower bound is the metallicity derived using the calibration of \citet{Pilyugin2005} (hereafter PT05)  and the upper bound using the calibration of \citet{Kobulnicky2004} (hereafter KK04).   As discussed in  \citet{Moustakas2010},  the PT05 calibration is based on empirical abundance measurements of individual HII regions and the KK04 calibration on photoionization model calculations.   The KK04 calibration gives abundances that are systematically higher than PT05.   In this paper we choose the PT05 calibration since this method was used to derive the XMPG abundances.  The results of this paper are unchanged if we use KK04 or an average of the two.

%%%%%%%%%%%%%%%%%%%%%%%%%%%%%%%%  table 11    %%%%%%%%%%%%%%%%%%%%%%%%%%%%%%%%%%%%%%%%%%%%%%%%%%%%%%%%%%%%%%%%%%%%%%
\begin{table}
 \caption{Characteristics of the SINGS sample.}
 \centering
 \small
 \begin{tabular}{l c c c c c c c c}
 \hline\hline
    Galaxy & Morphology & Distance & 12+log(O/H) &L(H$\alpha$)$_{obs}$  &  SFR  & N$_{ULX}$ &N$_{BKG}$\\ 
         &     &(Mpc) & &  (10$^{40}$ erg/sec) & (M$_{\odot}$/yr) &  \\ [0.5ex]
 \hline
 \multicolumn{7}{c}{{High-metallicity galaxies}}\\
 \hline
NGC 0925 & SAB(s)d & 9.12 & 8.24-8.78 & 6.01& 1.634 & 1& 0.566\\
NGC 2403 & SAB(s)cd & 3.5 & 8.31-8.81& 5.12 & 1.225 & 1& 0.474\\
NGC 2841 & SA(r)b & 9.8 & 8.52-9.19 & 3.39 & 3.17 & 2& 0.282\\
NGC 2976 & SAc & 3.5 & 8.30-8.98 & 0.87 & 0.334 & 0 & 0.028\\
NGC 3184 & SAB(rs)cd & 11.10 & 8.48-9.14 & 7.36 & 3.31 & 2&0.628 \\
NGC 3198 & SB(rs)c & 13.68 & 8.32-8.87 & 6.85 & 3.71& 1& 0.493\\
NGC 3351 & SB(r)b & 10.1 & 8.60-9.22  & 3.71 & 3.907 & 0 & 0.093 \\
NGC 3627 & SAB(s)b & 8.7 & 8.49-9.10 & 10.9 & 11.362 & 2 & 0.312\\
NGC 4559 & SAB(rs)cd & 11.1 & 8.25-8.79  & 10.6 & 2.545 & 3 & 0.578 \\
NGC 4569 & SAB(rs)ab & 16.58 & 8.56-9.19 & 4.84 & 6.731 & 1 & 1.015\\
NGC 4625 & SAB(rs)m & 9.17 & 8.27-9.04 & 0.61 & 0.215 & 0 & 0.013 \\
NGC 4736 & (R)SA(r)ab & 5.3 & 8.31-9.01 & 3.45 & 2.709 & 3 & 0.357 \\
NGC 4826 & (R)SA(rs)ab & 5.6 & 8.59-9.2 &4 4.05 & 2.732 & 0& 0.209\\
NGC 5055 & SA(rs)bc & 7.82 & 8.42-9.13 & 7.71 & 6.172 & 2 &0.613\\
NGC 5194 & SA(s)bc & 8.2 & 8.54-9.18 & 20.10 & 15.847 & 3&0.569 \\
NGC 6946 & SAB(rs)cd & 5.0 & 8.40-9.04 & 16.5 & 16.437 & 1& 0.4258\\
NGC 7331 & SA(s)b & 15.1 & 8.40-9.05 & 13.4 & 15.893 & 3 & 0.850\\
\hline
\multicolumn{7}{c}{{Intermediate-Metallicity Galaxies}}\\
\hline
NGC 1705 & SA0- & 5.1 & 8.20-8.43 & 0.9  &0.075 & 0 &0.009 \\
IC 2574 & SAB(s)m & 2.8 & 7.94-8.26  & 0.97 &0.128 & 0 & 0.084\\
NGC 4236 & SB(s)dm & 4.45 & 8.07-8.56 & 1.3 &0.183 & 0 & 0.411\\
IC 4710 & SB(s)m & 7.8 & 8.11-8.62 & 1.6 & 0.231 & 1&0.071\\
NGC 6822 & IB(s)m & 0.47 & 8.04-8.67 & 0.07 &0.015 & 0 &0.013\\
\hline
\multicolumn{7}{c}{{Low-Metallicity Galaxies}}\\
\hline
Ho II & Im & 3.5 & 7.68-8.07 & 0.64 & 0.073 & 1& 0.087 \\
DDO 053 & Im & 3.56 & 7.77-8.13 & 0.08 & 0.011 & 0 &  0.003\\
Ho IX & Im & 3.3 & 7.61-7.98  & 0.01 & 0.009 & 1& 0.008 \\
NGC 5408 & IB(s)m & 4.8 & 7.81-8.23 & 1.16 & 0.023 & 1 & 0.004\\
    \hline\hline \\
    \end{tabular} \\
\scriptsize
NOTES: Galaxy morphologies are from the NASA/IPAC Extragalactic Database (NED). Adopted distances as derived by \citet{Masters2005}. Oxygen abundances are from \citet{Moustakas2010} 
\label{tab:sings_sample}
\end{table} 
%%%%%%%%%%%%%%%%%%%%%%%%%%%%%%%%%%%%%%%%%%%%%%%%%%%%%%%%%%%%%%%%%%%%%%%%%%%%%%%%%%%%%%%%%%%%%%%%%%%%%%%%%%%%%%%%%%%%%

 We estimate the star formation rate of the SINGS galaxies using the calibration of \citet{Calzetti2010}:
 
\begin{equation}
  SFR(M_{\odot}/yr)=5.5\cdot10^{-42} \times [L(H\alpha)_{obs} + 0.02L(24\mu m)]
\end{equation}

 Note that the L(H$\alpha)_{obs}$ is the observed H$\alpha$ luminosity without correction for internal dust attenuation. 

 We use the measurements of the H$\alpha$ and 24$\mu$m fluxes of \citet{Dale2007} ) and \citet{Kennicutt2008} \& \citet{Kennicutt2009}.    The numbers of ULX  for the SINGS sample have been provided by \chandra\ SINGS team (Jenkins et al in preparation).  

\section{Comparison between the SINGS and XMPGs samples}
\label{nulx_compare}

\begin{figure}
  \centering
  \includegraphics{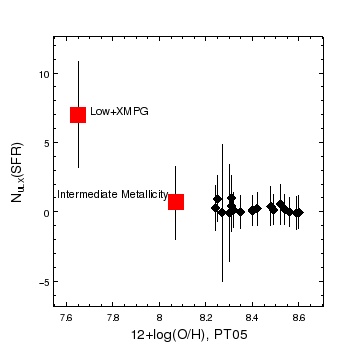}
  \caption{\nulx\  for individual SINGS galaxies, intermediate metallicity galaxies and the combined metal poor and XMPG.  This plot uses the PT05 metallicity calibration. }
  \label{nulx_norm}
\end{figure}

In this section, we investigate the relationship between metallicity and \nulx.    Table~\ref{tab:compare_ulx}  compares \nulx\ for the XMPG sample and the SINGS sample.  The  SINGS sample is divided into 3 sub-groups according to metallicity values (High, Intermediate and Low metallicity).   In addition, we show \nulx\ obtained by combining the SINGS low metallicity  sample and the XMPG sample.  The difference in \nulx\  between the high metallicity SINGS galaxies and the low metallicity galaxies (comprising the XMPG and low metallicity  SINGS galaxies) is 2.3$\sigma$.   The low metallicity sample has a small number of individual galaxies with very high N$_{ULX}$/SFR values (these have low SFR and one ULX).     A  Kolmogorov-Smirnov (K-S) test gives the probability that the two distributions (low metallicity and SINGS)  come from the same parent population as 0.18.     Finally, Figure~\ref{nulx_norm} shows \nulx\  as a function of metallicity.   The high metallicity SINGS galaxies are plotted as individual points,  and the XMPG and SINGS low metallictiy galaxies are combined.    We note that for galaxies with no ULX,  the background-subtracted number of ULX is negative, which is unphysical.   There is a marked increase in \nulx\ in the low metallicity galaxies.   We conclude that ULX form preferentially in low metal systems, with the caveat that the formal significance of this result is low.  

We do not find any any evidence for a trend in  \nulx\  with metallcity 12+log(O/H)$>$8.0.  Fitting the data points in Figure~\ref{nulx_norm}   above 12+log(O/H)$=$8.0 with both a flat line (slope 0) and a straight line with non-zero slope we find that the more complex model (non-zero slope) is slightly preferred on the basis of a $\chi^2$  fit.   We use an F-test to determine whether the model with a slope is significantly  better than the flat line (the null hypothesis is that a slope does not give a statistically better fit).    The F-test  gives a significance of 0.35,   confirming that a slope is not required.    

  \citet{Mapelli2010}  use a larger sample of galaxies and find an anti-correlation between \nulx\ and metallicity (their Figure 5 is directly comparable to Figure~\ref{nulx_norm} of this paper).     As discussed in the previous paragraph, there may be a similar trend in the SINGS galaxies but the effect is small below 12+log(O/H)$=$8.0 and the scatter is large.  We cannot unambiguously confirm this result.   The preference for ULX to form at the low metallicities is most apparent in the lowest metallicity bin.   

%%%%%%%%%%%%%%%%%%%%%%%%%%%%%%%%  table 15    %%%%%%%%%%%%%%%%%%%%%%%%%%%%%%%%%%%%%%%%%%%%%%%%%%%%%%%%%%%%%%%%%%%%%%
\begin{table}
 \caption{Comparison of the SINGS and XMPGs sample.}
 \centering
 \footnotesize
 \begin{tabular}{c c c c c c}
 \hline\hline
     & High & Intermediate &Low & XMPGs & Low$+$XMPG \\ 
 \hline
$\sum$N$_{ULX}$ & 25 &1 & 3& 5& 8 \\
$\sum$N$_{BKG}$ &7.5 & 0.6&  0.1 &1.3 & 1.4 \\
\nulx\  & 0.17$\pm$0.042&0.65$\pm$1.0 & 25.0$\pm$14.6& 4.5$\pm$2.3& 7.0$\pm$2.7 \\ 
\hline\hline \\
\end{tabular} 
\label{tab:compare_ulx}
\end{table}
%%%%%%%%%%%%%%%%%%%%%%%%%%%%%%%%%%%%%%%%%%%%%%%%%%%%%%%%%%%%%%%%%%%%%%%%%%%%%%%%%%%%%%

%----------------------------------- FIGURE 17-----------------------------------
 %------------------------------------------------------------------------------
 
 %----------------------------------- FIGURE 17-----------------------------------
\begin{figure}
  \centering
  \includegraphics{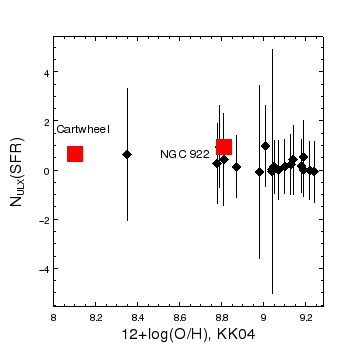}
  \caption{\nulx\  for individual SINGS galaxies, NGC 922 and the Cartwheel.  This plot uses the KK04 metallicity calibration. }
  \label{nulx_comp_cart_922}
\end{figure}
 %------------------------------------------------------------------------------

\subsection{Summary and Discussion}
\label{Conclusions}

In this paper, we present the results of a Chandra survey to search for ULX in the most metal poor galaxies.    We find that, compared to a comparison sample of high metallicity SINGS galaxies, the low metallicity galaxies are more likely to host a ULX.   The number of ULX normalized to the star formation rate is $\sim$ 0.17 for the high metallicity galaxies and 7.0 for the low metallicity systems (12+log(O/H)$<$8.0).   The formal significance of this result is low.   This study broadly agrees with the results of \citet{Mapelli2011} who also find that the number of ULX increases in lower metallicity galaxies.    \citet{Mapelli2011} also claim to see a trend in the numbers of ULX  as a function of metallicity.    As demonstrated in Figure~\ref{nulx_norm}, we do not see  strong evidence that  \nulx\ decreases as a function of metallicity  above $\sim 8.0$.      This result suggests there maybe a   "threshold" for more efficient ULX formation  at about   12+log(O/H)$<$8.

 In a recent paper, we  \citep{Prestwich2012}   compared the ULX population of two collisional ring galaxies, the Cartwheel  and NGC 922.     The Cartwheel has a relatively low  metallicity (12+log(O/H)$\sim$8.1) and NGC 922 has a metallicity close to solar (12+log(O/H)$\sim$8.81.   We  found that the number of ULXs in NGC 922 and the Cartwheel scales with the star formation rate: we do not find any evidence for an excess of sources in the Cartwheel relative to NGC 922.   The \nulx\ values for NGC 922 and the Cartwheel are the same (within the errors) as the SINGS galaxies.   This is demonstrated in Figure~\ref{nulx_comp_cart_922} where we plot \nulx\ vs. metallicity for the SINGS galaxies, the Cartwheel and NGC 922.   The KK04 calibration is used for this plot as it is more appropriate for the Cartwheel and NGC 922 measurements.     Figure~\ref{nulx_comp_cart_922} adds to the evidence that the high number of ULX in the Cartwheel is due to it's very high star formation rate, and not a metallicity effect.   
 
There are two hypotheses which might explain the excess of ULX in metal-poor galaxies.   \citet{Linden2010} studied the effect of metallicity on HMXB production and found that the number of ULX in low metal systems increased dramatically below a threshold of Z/Z$_{\odot} <$10\%.     The high X-ray luminosities derive from Roche Lobe Overflow onto a black hole, which typically has a mass M$\le 10$M$_{\odot}$.    The  threshold  predicted by  \citet{Linden2010}  is consistent with the increase in \nulx\ we see in galaxies with 12+log(O/H)$<$8.    An alternate hypothesis is that higher mass black holes are able to form in lower metallicity gas \citep{Mapelli2009, Zampieri2009} leading to higher X-ray luminosities in HMXB.     Our results are consistent with both of these scenarios.   

\section{Acknowledgements}

 Thanks to the Leigh Jenkins and the \chandra\ SINGS teams for providing results prior to publication, to  Doug Swartz for giving us details of galaxies with no detected ULX and to the referee Michela Mapelli for helping to clarify many points in the paper.    Support for this work was provided by the National Aeronautics and
Space Administration through Chandra Award Number G0-11108A issued by the
Chandra X-ray Observatory Center, which is operated by the Smithsonian
Astrophysical Observatory for and on behalf of the National
Aeronautics Space Administration under contract NAS8-03060.   This research has made use of the NASA/IPAC Extragalactic Database (NED) which is operated by the Jet Propulsion Laboratory, California Institute of Technology, under contract with the National Aeronautics and Space Administration.   Based on observations made with the NASA Galaxy Evolution Explorer. 
GALEX is operated for NASA by the California Institute of Technology under NASA contract NAS5-98034

%%%%%%%%%%%%%%%%%%%%%%%%%%%%%%%%%%%%%% SECTION 5 %%%%%%%%%%%%%%%%%%%%%%%%%%%%%%%%%%%%%%%%%%%%%%%%%%%%%%%%%%%%%%%

\addcontentsline{toc}{section}{References}
%\bibliographystyle{plain}

%%%%%%%%%%%%%%%%%%%%%%%%%%%%%%%%%%%%%%%%%%%%%%%%%%%%%%%%%%%%%%%%%%%%%%%%%%%%%%%%%%%%%%%%%%%%%%%%%%%%%%%%%%%%%%%%%


\begin{thebibliography}{}


\bibitem[Calzetti et al.(2007)]{Calzetti2007} Calzetti, D., et al.\ 2007, 
\apj, 666, 870 


\bibitem[Calzetti et al.(2010)]{Calzetti2010} Calzetti, D., et al.\ 2010, 
\apj, 714, 1256 


\bibitem[Colbert 
\& Mushotzky (1999)]{Colbert1999} Colbert, E.~J.~M., \& Mushotzky, R.~F.\ 1999, \apj, 519, 89 



\bibitem[Dale et al. (2007)]{Dale2007} Dale, D.~A., et al.\ 2007, \apj, 655, 
863 


\bibitem[de Vaucouleurs et al.(1991)]{deVaucouleurs1991} de Vaucouleurs, 
G., de Vaucouleurs, A., Corwin, H.~G., Jr., Buta, R.~J., Paturel, G., 
\& Fouque, P.\ 1991, Volume 1-3, XII, 2069 pp.~7 figs..~ Springer-Verlag Berlin Heidelberg New York,  


\bibitem[Dickey \& Lockman(1990)]{Dickey1990} Dickey, J.~M., \& Lockman, F.~J.\ 1990, \araa, 28, 215 


\bibitem[Farrell et al.(2009)]{Farrell2009} Farrell, S.~A., Webb, N.~A., 
Barret, D., Godet, O., \& Rodrigues, J.~M.\ 2009, \nat, 460, 73 


\bibitem[Feng \& Soria(2011)]{Feng2011} Feng, H., \& Soria, R.\ 2011, \nar, 55, 166 


\bibitem[Freeman et al. (2002)]{Freeman2002} Freeman, P.~E., Kashyap, V., Rosner, R., \& Lamb, D.~Q.\ 2002, \apjs, 138, 185 


\bibitem[Gehrels(1986)]{Gehrels1986} Gehrels, N.\ 1986, \apj, 303, 336 

\bibitem[Giacconi et al.(2001)]{Giacconi2001} Giacconi, R., et al.\ 2001, 
\apj, 551, 624 

\bibitem[Grimm et al. (2003)]{Grimm2003} Grimm, H.-J., Gilfanov, M., \& Sunyaev, R.\ 2003, \mnras, 339, 793 

§
\bibitem[Gilfanov, Grimm, 
\& Sunyaev(2004)]{Gilfanov2004_2} Gilfanov, M., Grimm, H.-J., \& Sunyaev, R.\ 2004, \mnras, 347, L57 


\bibitem[Gladstone at al. (2009)]{Gladstone2009} Gladstone, J.~C., Roberts, T.~P., \& Done, C.\ 2009, \mnras, 397, 1836 


\bibitem[Goad et al. (2006)]{Goad2006} Goad, M.~R., Roberts, T.~P., Reeves, J.~N., \& Uttley, P.\ 2006, \mnras, 365, 191 

\bibitem[Hunter et al. (2010)]{Hunter2010} Hunter, D.~A., Elmegreen, B.~G., \& Ludka, B.~C.\ 2010, \aj, 139, 447 

\bibitem[Kaaret, Schmitt, 
\& Gorski(2011)]{Kaaret2011} Kaaret, P., Schmitt, J., \& Gorski, M.\ 2011, \apj, 741, 10 

\bibitem[Kennicutt et al.(2003)]{Kennicutt2003} Kennicutt, R.~C., Jr., et 
al.\ 2003, \pasp, 115, 928 


\bibitem[Kennicutt et al.(2009)]{Kennicutt2009} Kennicutt, R.~C., Jr., et 
al.\ 2009, \apj, 703, 1672 


\bibitem[Kennicutt et al.(2008)]{Kennicutt2008} Kennicutt, R.~C., Jr., Lee, 
J.~C., Funes, S.~J., Jos{\'e} G., Sakai, S., 
\& Akiyama, S.\ 2008, \apjs, 178, 247 

\bibitem[Kobulnicky 
\& Kewley(2004)]{Kobulnicky2004} Kobulnicky, H.~A., \& Kewley, L.~J.\ 2004, \apj, 617, 240 



\bibitem[Linden et al.(2010)]{Linden2010} Linden, T., Kalogera, V., 
Sepinsky, J.~F., Prestwich, A., Zezas, A., 
\& Gallagher, J.~S.\ 2010, \apj, 725, 1984 


\bibitem[Liu et al. (2007)]{Liu2007} Liu, J.-F., Bregman, J., Miller, J., \& Kaaret, P.\ 2007, \apj, 661, 165 


\bibitem[Liu 
\& Bregman(2005)]{Liu2005} Liu, J.-F., \& Bregman, J.~N.\ 2005, \apjs, 157, 59 

\bibitem[Masters (2005)]{Masters2005} Masters, K., 2005, PhD Thesis, Cornell University

\bibitem[Mapelli et al. (2009)]{Mapelli2009} Mapelli, M., Colpi, M., \& Zampieri, L.\ 2009, \mnras, 395, L71 


\bibitem[Mapelli et al. (2011)]{Mapelli2011} Mapelli, M., Ripamonti, E., Zampieri, L., \& Colpi, M.\ 2011, Astronomische Nachrichten, 332, 414 


\bibitem[Mapelli et al.(2010)]{Mapelli2010} Mapelli, M., Ripamonti, E., 
Zampieri, L., Colpi, M., \& Bressan, A.\ 2010, \mnras, 408, 234 


\bibitem[Miller 
\& Colbert(2004)]{Miller2004} Miller, M.~C., \& Colbert, E.~J.~M.\ 2004, International Journal of Modern Physics D, 13, 1 


\bibitem[Mineo et al. (2012)]{Mineo2012} Mineo, S., Gilfanov, M., \& Sunyaev, R.\ 2012, \mnras, 419, 2095 

\bibitem[Moustakas et al.(2010)]{Moustakas2010} Moustakas, J., Kennicutt,  R.~C., Jr., Tremonti, C.~A., Dale, D.~A., Smith, J.-D.~T.,  \& Calzetti, D.\ 2010, \apjs, 190, 233 


\bibitem[Papaderos et al. (2008)]{Papaderos2008} Papaderos, P., Guseva, N.~G., Izotov, Y.~I., \& Fricke, K.~J.\ 2008, \aap, 491, 113 

\bibitem[Pilyugin \& Thuan(2005)]{Pilyugin2005} Pilyugin, L.~S., \& Thuan, T.~X.\ 2005, \apj, 631, 231 


\bibitem[Prestwich et al.(2007)]{Prestwich2007} Prestwich, A.~H., et al.\ 
2007, \apjl, 669, L21 

\bibitem[Prestwich et al.(2012)]{Prestwich2012} Prestwich, A.~H., et al.\ 
2012, \apj, 747, 150 

\bibitem[Ranalli, Comastri, 
\& Setti(2003)]{Ranalli2003} Ranalli, P., Comastri, A., \& Setti, G.\ 2003, \aap, 399, 39 



\bibitem[Roberts et al.(2006)]{Roberts2006} Roberts, T.~P., Kilgard, R.~E., 
Warwick, R.~S., Goad, M.~R., \& Ward, M.~J.\ 2006, \mnras, 371, 1877 


\bibitem[Roberts (2007)]{Roberts2007} Roberts, T.~P.\ 2007, \apss, 311, 203 

\bibitem[Soria (2007)]{Soria2007} Soria, R.\ 2007, IAU Symposium, 238, 235 


\bibitem[Soria et al.(2005)]{Soria2005} Soria, R., Cropper, M., Pakull, M., 
Mushotzky, R., \& Wu, K.\ 2005, \mnras, 356, 12 


\bibitem[Sutton et al.(2012)]{Sutton2012} Sutton, A.~D., Roberts, T.~P., 
Walton, D.~J., Gladstone, J.~C., \& Scott, A.~E.\ 2012, \mnras, 423, 1154.


\bibitem[Swartz et al.(2004)]{Swartz2004} Swartz, D.~A., Ghosh, K.~K., Tennant, A.~F., \& Wu, K.\ 2004, \apjs, 154, 519 


\bibitem[Swartz et al. (2008)]{Swartz2008} Swartz, D.~A., Soria, R., \& Tennant, A.~F.\ 2008, \apj, 684, 282 

\bibitem[Swartz et al. (2011)]{Swartz2011} Swartz, D.~A., Soria, R., Tennant, A.~F., \& Yukita, M.\ 2011, \apj, 741, 49 


\bibitem[Thuan et al. (2004)]{Thuan2004} Thuan, T.~X., Bauer, F.~E., Papaderos, P., \& Izotov, Y.~I.\ 2004, \apj, 606, 213 

\bibitem[Walton et al. (2011)]{Walton2011} Walton, D.~J., Roberts, T.~P., Mateos, S., \& Heard, V.\ 2011, \mnras, 416, 1844 


\bibitem[Zampieri 
\& Roberts(2009)]{Zampieri2009} Zampieri, L., \& Roberts, T.~P.\ 2009, \mnras, 400, 677 


\bibitem[Zezas et al.(2007)]{Zezas2007} Zezas, A., Fabbiano, G., Baldi, A., 
Schweizer, F., King, A.~R., Rots, A.~H., 
\& Ponman, T.~J.\ 2007, \apj, 661, 135 

\end{thebibliography}
\end{document}